\documentclass[a4paper,11pt]{article}
\usepackage {epsfig}
\usepackage[hmargin=3cm,vmargin=3cm]{geometry}
\usepackage{amsmath}

\usepackage{amsmath,amsthm,amsfonts,amssymb}
\usepackage{graphicx}
\usepackage{epstopdf}
\usepackage{bm}
\usepackage{natbib}
\usepackage{color}

\newcommand{\eps}{{\varepsilon}}
\newcommand{\Lop}{{\mathcal{L}}}
\newcommand{\LOop}{{\mathcal{L}}_0}
\newcommand{\x}{{\bm{\mathit{x}}}}
\newcommand{\y}{{\bm{\mathit{y}}}}
\newcommand{\z}{{\bm{\mathit{z}}}}
\newcommand{\zi}{{\bm{\mathrm{z}}}}
\newcommand{\Proj}{{\mathbf{\mathcal P}}}
\newcommand{\Q}{{\mathbf{\mathcal Q}}}
\newcommand{\Id}{{\mathbf{\mathcal I}}}
\newcommand{\A}{{\mathbf{\mathcal A}}}
\newcommand{\B}{{\mathbf{\mathcal B}}}
\newcommand{\R}{{\mathbb R}}
\newcommand{\EE}{{\mathbb E}}

\def\Lam{{\bf{\Lambda}}}

\begin{document}

\title{\bf \Large Stochastic Climate Theory}
\author{Georg A. Gottwald$^{1}$, Daan T. Crommelin$^{2,3}$, and Christian L. E. Franzke$^{4}$}

\date{}

\maketitle

\begin{center}
{\small
\noindent $^1$ School of Mathematics and Statistics, The University of
Sydney, Sydney, Australia\\
\noindent $^2$ CWI, Amsterdam, The Netherlands\\
\noindent $^3$ Korteweg-de Vries Institute for
Mathematics, University of Amsterdam, The Netherlands\\
\noindent $^4$ Meteorological Institute and Center for Earth System
Research and Sustainability, University of Hamburg, Hamburg,
Germany
 }
\end{center}

%%%%%%%%%%%%%%%%%%%%%%%%%%%%%%%%%%%%%%%%%%%%%

\begin{abstract}
In this chapter we review stochastic modelling methods in climate science. First we provide a conceptual framework for stochastic modelling of deterministic dynamical systems based on the Mori-Zwanzig formalism. The Mori-Zwanzig equations contain a Markov term, a memory term and a term suggestive of stochastic noise. Within this framework we express standard model reduction methods such as averaging and homogenization which eliminate the memory term. We further discuss ways to deal with the memory term and how the type of noise depends on the underlying deterministic chaotic system. Secondly, we review current approaches in stochastic data-driven models. We discuss how the drift and diffusion coefficients of models in the form of stochastic differential equations can be estimated from observational data. We pay attention to situations where the data stems from multi scale systems, a relevant topic in the context of data from the climate system. Furthermore, we discuss the use of discrete stochastic processes (Markov chains) for e.g. stochastic subgrid-scale modeling and other topics in climate science.
\end{abstract}

\section{Introduction}
\label{sec-intro}
The climate system is characterized by the mutual interaction of
complex systems each involving entangled processes running on spatial
scales from millimeters to thousands of kilometers, and temporal
scales from seconds to millennia. Given current computer power it is
impossible to capture the whole range of spatial and temporal scales
and this will also not be possible in the foreseeable
future. Depending on the question we pose to the climate system - be
it forecasting regimes in the atmosphere or simulating past ice ages - we
have to make a decision as to what components to include in the
analysis and as to what scales to resolve. A corollary of this
decision is that each numerical scheme inevitably fails to resolve so
called \emph{unresolved scales} or \emph{subgrid-scales}. However, typically
one is only interested at the slow processes active on large spatial
scales. For example, for weather forecasts it is sufficient to resolve large
scale high and low pressure fields rather than small scale fast
oscillations of the stratification surfaces, whereas for climate
predictions with a coupled ocean-atmosphere model we may want to distill
the slow dynamics of the ocean ignoring weather systems interacting
with the ocean on fast time-scales of days.

The dynamics of the unresolved scales, however, have a significant
impact on the large scales and simply ignoring them has detrimental
effects on reliably simulating the slow large scale variables of
interest. For example, \cite{DawsonPalmer14} showed that the ECMWF model
produces unrealistic spatial patterns of atmospheric weather regimes due to
not sufficiently resolving the small-scale processes. This has implications
for the current global circulation models used for the intergovernmental Panel
on Climate Change fifth assessment report (IPCC AR5) which typically use
coarser resolutions.

The question we are concerned with in this chapter is whether it is
possible to obtain reliable simulations of the slow large-scale
characteristics of the climate system without having to resolve the
small fast scales accurately by employing computationally very costly
high-resolution simulations but rather by parametrizing them by judiciously
chosen noise, and if so, under what conditions? Heuristically this
should be possible in the following situations \citep{Givonetal04}:
1.) time scale separation and 2.) weak coupling to a large system.

In a time-scale
separated system, during one slow-time unit the fast uninteresting
variables $y$ perform many ``uncorrelated" events (provided the fast
dynamics is sufficiently chaotic). The contribution of the
uncorrelated events to the dynamics of the slow interesting variable
$x$ is as then a sum (or integral) of independent random variables. By the Central Limit
Theorem this can be expressed by a normally distributed variable. Similarly, if a large number of uninteresting variables $y$ are weakly
coupled to the resolved interesting variables $x$, it takes many
uncorrelated events of the unresolved variables to have a significant
effect on the dynamics of the resolved variables. The resolved
variables $x$ experience a cumulative contribution of those events,
which again by the Central Limit Theorem allows us to parameterise the
unresolved ``heat bath" $y$ by a random process. Here the randomness
is not mediated by chaotic dynamics and time-scale separation, but by
a large number of weakly coupled variables with random initial
conditions drawn typically from some thermodynamic equilibrium
density.

Note that in both cases, the stochasticity arises only
asymptotically, by either infinite time scale separation or by an
infinitely large heat bath. Real life applications never satisfy these
limits and care has to be taken. \citet{DelSole00} pointed out, for
example, that on short time scales stochastic models are not able to
capture deterministic dynamics. Short meaning here that the fast
chaotic variables have not sufficiently decorrelated to allow for the
central limit theorem to act.

In the climate science community it has been realized that
stochasticity may be used to parametrize subgrid-scale phenomena. In
climate modeling, the idea of modeling fast chaotic dynamics by
stochastic processes and thereby reducing the effective dimension of
the full system goes back to the seminal works by \cite{Hasselmann76}
and \cite{Leith75}. In their work \cite{Hasselmann76} and
\cite{Leith75} have suggested studying climatic regime switches by
introducing in an ad-hoc way a stochastic driver for the slow
dynamics. Such an approximation describes the deviations from an
averaged climatological system. Of course, it is natural to expect
such behavior only if the fast variables (eg. weather in a coupled
climatic ocean-atmosphere model) are sufficiently chaotic.

These ideas have been used to simulate systems of increasing
complexity including the barotropic vorticity equation
\citep{DuanNadiga07,Franzke05}, a 3-layer quasi-geostrophic
prototype climate model \citep{FranzkeMajda06} and a primitive equation
model \citep{ZhangHeld99}.

The effective dimension reduction achieved if a large number of fast
equations are replaced by only a few stochastic process, and the
associated computational advantage of such a reduction is a huge
driving force behind this research. Such reduction strategies also
provide insight into the underlying dynamics of the climate system and
pose new challenging mathematical problems.

The ``Hasselmann program'', as coined by \citet{Arnold01}, of
stochastic model reduction, which has received renewed attention in
the past few years, has not been finished yet, and poses a fascinating
challenge for mathematicians. In particular, how can we make the
transition from a purely deterministic system to a stochastic system
in a controllable way. In the following we will introduce a formalism
which allows us to rewrite a deterministic system in a way that
``looks like" a stochastic system in the form of generalized Langevin
equations, and which may be a formal starting point for controlled
stochastic model reduction.

%%%%%%%%%%%%%%%%%%%%%%%%%%%%%%%%%%%%%%%%%%%%%

\section{Conceptual framework for stochastic modeling: The
  Mori-Zwanzig formalism}
In a series of seminal papers \citet{Mori65a,Mori65b} and
\citet{Zwanzig73} developed a formalism to rewrite a deterministic
dynamical system in a form which resembles a general Langevin equation
of the form
\begin{align}
\frac{dz}{dt} = f(z(t)) + \int_0^tK(z(t-s),s) ds + \dot W_t \; .
\label{e.GLE}
\end{align}
The first term is Markovian, the second term describes possible memory
of the process and $W_t$ denotes a stochastic process. The
Mori-Zwanzig formalism provides a conceptual framework to study
dimension reduction and to parametrize uninteresting variables by a
stochastic process.

We first briefly review the Mori-Zwanzig formalism before
formulating standard deterministic and stochastic parameterization
techniques such as averaging and homogenization within this
framework.

%%%%%%%%%%%%%

\subsection{The Mori-Zwanzig projection operator formalism}
\label{sec-son}
The main idea behind the reformulation of deterministic dynamics is
simple and can be understood by the method of variation of
constants. The following illustrative example is taken from the book
by \cite{Zwanzig}. Consider the coupled linear system
\begin{align*}
\dot{x}&= L_{11}x+L_{12}y\\
\dot{y}&= L_{21}x+L_{22}y\; .
\end{align*}
Suppose we are only interested in the dynamics of $x$, and only have
climatic knowledge of the initial condition of the variable $y$,
i.e. its mean and variance. We can then solve for $y$ to obtain
\[
y(t)=e^{L_{22}t}y(0)+\int_0^te^{L_{22}(t-s)}L_{21}x(s)\, ds\; ,
\]
which we may use to express the dynamics of the ``interesting" variable as
\[
{\dot{x}} = L_{11} x + L_{12}\int_0^te^{L_{22}(t-s)}L_{21}x(s)\, ds + L_{12}e^{L_{22}t}y(0)\; .
\]
This is of the form of a generalized Langevin equation (\ref{e.GLE}),
where the first term is Markovian, the second term a memory term, and
the third term is a noise term if we treat the initial conditions $y(0)$ as noise.
%randomly distributed the last term can be interpreted as a noise term.

Let us now consider the general nonlinear case. Consider the
deterministic dynamical system
\begin{align}
\dot{\zi} = \mathbf f(\zi)\; ,
\label{e.ds}
\end{align}
with initial condition $\zi(0)=\zi_0$. Here $\z$ is either a
finite-dimensional state vector $\zi\in \R^d$ or an element of a
Hilbert space. Associated with the typically nonlinear dynamical
system (\ref{e.ds}) is the following linear partial differential
equation for the temporal evolution of an observable $v(\z,t)$
\begin{align}
\frac{\partial v}{\partial t}=\Lop v\quad {\rm with} \quad
v(\z,0)=\Phi(\z)\; ,
\label{eq.cp}
\end{align}
with the generator
\begin{align}
\Lop = \mathbf f(\z)\cdot \nabla\, ,
\label{e.gen}
\end{align}
where $ \nabla $ denotes the gradient in phase space, i.e. $\mathbf f(\z) \cdot \nabla = f_i(\z)\partial_{z_i}$.
The solution of (\ref{eq.cp}) is formally written as
\begin{align}
v(\z,t) = e^{\Lop t}\Phi(\z)\; .
\label{eq.expL}
\end{align}
The equivalence of the nonlinear ordinary differential equation
(\ref{e.ds}) and the linear partial differential equation
(\ref{eq.cp}) can be seen mathematically by employing the chain rule
on $v(\zi(t))$ or by the following heuristic consideration. To
determine the value of an observable at time $t$ one may either follow
a trajectory and evaluate the observable along the trajectory or one
may follow the evolution of the actual observable along the
characteristic, i.e. $v(\z,t) = \Phi(\zi(t))$ with $\zi(0)=\z$. We
assume here that the vector field $\mathbf f(\z)$ is smooth enough to
assure uniqueness and existence of solutions of (\ref{e.ds}) and of
classical solutions of the associated partial differential equation
(\ref{eq.cp}). Note that $\Lop$ is the formal $L^2$-adjoint operator
of the Liouville operator $\Lop^\star$ with $\Lop^\star \rho = -\nabla
\cdot(f(\z)\rho)$, controlling the evolution of densities of ensembles
propagated according to (\ref{e.ds}).

Suppose one is not interested in resolving the full solution $\z(t)$
but rather is interested in only a few observables
$\Phi(\z)=\left(\Phi_1(\z), \Phi_2(\z), \cdots,\Phi_n(\z)\right)$. A
particular relevant example occurs in the situation where the state
space can be decomposed as $\z=(\x,\y)$ into ``interesting variables"
$\x=(z_1,\cdots,z_n)\in\R^n$ and the remainder of ``uninteresting"
variables $\y=(z_{n+1},\cdots,z_d )$. The resolved observables would be
$\Phi(\z)=(z_1,\cdots,z_n)$ in this case. In the infinite dimensional
case where (\ref{e.ds}) denotes a partial differential equation, one may consider the
dynamical system (\ref{e.ds}) as a Galerkin approximation and  the resolved observables
could, for example, be the low-order Fourier modes of a spectral
representation of the solution, i.e.
the relevant variables are those with small wavenumbers $k_< = \{ k:
k\le k^\star \}$ and the irrelevant variables as those with high
wavenumbers $k_> = \{ k: k>k^\star \}$.
The question we
are concerned with in model reduction is how to distill the effective
dynamics of these ``interesting" observables?

To distill the dynamics of the interesting variables $\Phi(\z)$ we
require a projection operator $\Proj$ which maps functions of $\z$ to
functions of $\Phi(\z)$. In the
context of partial differential equations, the projection operator can
then be defined, for example, as $(\Proj \omega({\bf k}))({\bf k}_<) = \omega({\bf k}_<,0)$ for functions $\omega({\bf k})$. Let us restrict for simplicity of exposition
to the case where $\z=(\x,\y)$ and $\Phi(\z)=\x\in \R^n$. A suitable projector for the situation when the initial conditions of
the interesting variables $\x$ are known exactly but only statistical
information is available for the unresolved variables $\y$, is the
conditional expectation of a function $\omega(\x,\y)$
\begin{align}
\label{e.P}
(\Proj \omega)(\x)= \frac{1}{\Omega(\x)} \int_{\R^{d-n}}\omega(\bm
\x,\bm \y)\, \mu_\x(d\y)\; ,
\end{align}
where $\mu_\x(d\y)$ denotes the conditional measure of the unresolved
variables. The normalization
\[
\Omega(\x)=\int_{\R^{d-n}}\mu_{\x}(d \y)
\]
measures in the language of statistical mechanics the number of
\emph{microstates} which give rise to the \emph{macrostate} $\x$.

In the context of equilibrium statistical mechanics  often a unique
invariant measure exists  which supports a density with respect to the
Lebesgue measure with $\mu_\x(d\y) = \rho_{\rm eq}(\y|\x) d\y$. In the
context of deterministic dynamical systems typically a multitude of
ergodic measures exist and the value of $(\Proj \omega)(\x)$ would
depend on the choice of the initial conditions of $\y$. To complicate
things further, these measures may not depend continuously on $\x$
(see below). These measures are singular and their support is not on a
set of full Lebesgue-measure but rather on an attractor or on a
surface of constant energy in the case of dissipative and conservative
deterministic dynamical systems, respectively. Nevertheless, for a
large class of dynamical systems one can introduce the notion of a
physical measure which supports densities on the surfaces of constant
energy or on the attractor. In the case of (dissipative) chaotic
deterministic dynamical systems these are given by so called
Sinai-Ruelle-Bowen (SRB) measures \citep{Young98,Young99,Young02}. SRB
measures $\mu^{\rm SRB}$ satisfy the property that for a set of
non-zero Lebesgue measure initial conditions $\z(0)$ and for every
continuous observable $\varphi$ we have
\begin{align*}
\lim_{T\to \infty}\frac{1}{T}\int_0^T\varphi(\z(t)) dt \to \int \varphi \, \mu^{\rm SRB}\; .
\end{align*}
This property assures that meaningful averages can be calculated and
the statistics of the dynamical system can be explored by the
asymptotic distribution of orbits starting from Lebesgue almost every
initial condition. The class of systems for which SRB are proven to
exist includes for example, uniformly hyperbolic systems, logistic-map
type systems, H\'enon-like attractors, Lorenz attractors and many
more. It has recently been conjectured by \cite{GottwaldMelbourne14}
that typical dynamical systems are either regular or belong to the
above class which enjoys good statistical properties. In the following
all measures are understood to be SRB measures. Furthermore we assume
that all measures are normalized to $\int\mu = 1$.

Given a projection operator $\Proj$, we denote by $\Q=\bm 1 -\Proj$
the orthogonal projector. We then write the problem
(\ref{eq.cp}) as
\begin{align*}
\frac{\partial v}{\partial t}(\z,t) &=\Lop e^{\Lop t}\Phi(\z)\\
&= e^{\Lop t}\Proj\Lop \Phi(\z)+e^{\Lop t}\Q\Lop \Phi(\z)\; ,
\end{align*}
which upon using the Duhamel-Dyson formula (see, for example,
\cite{EvansMorris}) for operators $\A$ and $\B$
\begin{align}
\label{eq.Duhamel}
e^{t(\A+\B)} = e^{t\A} + \int_0^t e^{(t-s)(\A+\B)} \, \B \, e^{s \A}\, ds\; ,
\end{align}
becomes the celebrated Mori-Zwanzig equation \citep{Mori65a,Zwanzig73}
\begin{align}
\frac{\partial v}{\partial t}(\z,t)=
e^{\Lop t}\Proj\Lop \Phi(\z)
+ \int_0^t e^{(t-s)\Lop} \, \Proj\Lop \, e^{s \Q\Lop}\Q\Lop\Phi(\z)\, ds
+ e^{t\Q\Lop} \Q\Lop \Phi(\z) \; ,
\label{eq.MZ}
\end{align}
with $ \A = \Q\Lop $, $ \B = \Proj\Lop $ and $ \A + \B = \Lop $.

The Mori-Zwanzig equation (\ref{eq.MZ}) is not an approximation but is
exact and constitutes an equivalent formulation of the full dynamical
system (\ref{e.ds}). The interested reader is referred to
\cite{Zwanzig,ChorinHald,EvansMorris,Chorin00,Givonetal04} for more
details.\\

The Mori-Zwanzig equation (\ref{eq.MZ}) is in the form of a
generalized Langevin equation: the first term on the right-hand side
is Markovian while the second term involves memory. The last term
$n(\z,t) = e^{t\Q\Lop} \Q\Lop \Phi(\z) $ is labeled the noise
term. This is because its temporal evolution
\begin{align}
\frac{\partial n}{\partial t}(\z,t)=\Q\Lop n(\z,t)\quad {\rm with} \quad
n(\z,0)=\Q\Lop\phi(\z)\; ,
\end{align}
assures that the dynamics remains orthogonal to the range of $\Proj$. In the
case $\Phi(\z)=\x$ where we split the dynamical system (\ref{e.ds})
into the resolved and unresolved variables $\x$ and $\y$ respectively,
as
\begin{align}
\dot \x &= f(\x,\y)\\
\dot \y &= g(\x,\y)\;,
\end{align}
the orthogonal dynamics describes the dynamics of the fluctuating part
of the vector field of the resolved variables since
$n(\z,0)=f(\x,\y)-(\Proj f)(\x)$.

It is a formidable challenge to find effective approximations which
render the Mori-Zwanzig equation as a closed equation for the resolved
variables and finding approximations for the noise term and the
infinite-dimensional memory kernel. Obviously this programme can only
be exercised within approximations. In the following section we will
describe a simple formal procedure to unravel the memory kernel.\\

%%%%%%%%%%%%%

\subsection{Writing the memory term as an infinite chain of Markov terms}
\label{e.MZ-snake}
Several approximations have been employed to simplify the memory
term. \citet{Mori65a} considered the case where the dynamics is given
by a Hamiltonian system and the projection operator is the average
over all variables. In this case the Mori-Zwanzig equation is a linear
equation for the resolved variables and the Laplace transform of the
memory kernel could be written as a continued fraction rendering the
Mori-Zwanzig equation as a Markov chain. An extension to the
non-Hermitian case was given by \cite{Grigolini82}. In the nonlinear
case the short-memory approximation was introduced \citep{Chorin00}
which allows for an analytical treatment. Loosely speaking this
assumption states that the resolved and the unresolved subspaces do
not couple, and one may use the full dynamics to propagate
the elements of the orthogonal subspace. In the short-memory
approximation the memory term is replaced by a damping term which is linear in
the time variable $t$ which renders these approximation unsatisfactory
for long time integrations and for estimating the statistics of the
slow variables.

We will now present a simple reformulation of the Mori-Zwanzig
equation which allows for a computationally accessible criterion for a
truncation of the memory term. We rewrite problem (\ref{eq.cp}) as
\begin{align*}
\frac{\partial v}{\partial t}(\z,t)
&= e^{\Lop t}\Proj\Lop \Phi(\z)+e^{\Lop t}\Q\Lop \Phi(\z)\; .
\end{align*}
The second term, as we have seen in the previous section, can be
written as the sum of a memory kernel and a noise-like term, here
however, we express the second term as a time dependent forcing of the
first Markovian term, and specify its time evolution
\begin{align*}
\frac{\partial v}{\partial t}(\z,t)
&= e^{\Lop t}\Proj\Lop \Phi(\z)+n_1(t)\\
\frac{\partial n_1}{\partial t}(\z,t)
&= \Lop n_1(t)
\; ,
\end{align*}
with
\[
n_1(t)= e^{\Lop t}\Q\Lop \Phi(\z)\;.
\]
Note that $n_1$ and $v$ solve the same linear partial differential equation.
Repeating this process for the equation for $n_1$ we arrive at the
infinite Markov chain
\begin{align*}
\frac{\partial v}{\partial t}(\z,t)
&= \Lam_0 \Phi(\z)+n_1(\z,t)\\
\frac{\partial n_1}{\partial t}(\z,t)
&= \Lam_1 n_1(\z) +n_2(\z,t)\\
&\;\;\vdots\\
\frac{\partial n_k}{\partial t}(\z,t)
&=  \Lam_k n_k(\z)+n_{k+1}(\z,t)\\
&\;\; \vdots
\end{align*}
with the operator of the Markov term
\begin{align}
\Lam_k = e^{\Lop t}\Proj_k\Lop
\label{e.MZLambda}
\end{align}
and the forcing
\[
n_k(\z,t)= e^{\Lop t}\Q_k\Lop n_{k-1}(\z)\; ,
\]
with $n_0(\z)=\Phi(\z)$. Note that we allow for different projectors
$\Proj_k$ and $\Q_k=\bm 1 -\Proj_k$ at different levels. The advantage
of this formulation is that we can introduce a condition to
truncate this infinite series, thereby approximating the
infinite-dimensional memory term. The Markov chain can be truncated at
level $k$ provided the autocorrelation function of the remainder
$\langle n_{k+1}(t)n_{k+1}(s) \rangle$ corresponds to some known noise
process, for example, if
\begin{align}
\label{e.trunc1}
\langle n_{k+1}(t)n_{k+1}(s) \rangle = \sigma^2 \delta(t-s)\; .
\end{align}
Using the Mori-Zwanzig formalism to unravel possible memory and
long-time persistence of the dynamics by enlarging the state-space
until closure can be obtained has been algorithmically realized in the
multi-level regression ideas promoted in \cite{Kravtsov05} (see also
\cite{Chekroun11}). It has been employed, for example, to model the
El-Ni\~no-Southern Oscillation system \citep{KondrashovEtAl05} and
low-frequency variability in a three-level quasi-geostrophic model
\citep{KondrashovEtAl06}. Their particular implementation of restricting
the vector fields $\Lambda_k$ in (\ref{e.MZLambda}) to quadratic
polynomials in those works was pointed out to lead to undesirable
instabilities in energy-conserving systems by
\cite{MajdaYuan12}. Attempts to remedy those short-comings whilst
keeping the general idea of \cite{Kravtsov05}, and thereby of the
framework advocated here, were proposed by
\cite{MajdaHarlim13} and \citet{KondrashovEtAl15}.

In the following we discuss two generic situations for which
truncations of the Markov chain can be rigorously justified and for
which the memory kernel vanishes all together. At the end of this section we briefly describe some promising new directions in going beyond the assumption of infinite-timescale separation underlying the rigorous theory.

%%%%%%%%%%%%%

\subsection{Averaging in the Mori-Zwanzig framework}
\label{sec.av}
We will consider systems for slow variables $x\in\R^n$ and fast
variables $y\in\R^m$ of the form
\begin{align}
\dot x &=f_0(x,y)
\label{e.av1}
\\
\dot y &= \frac{1}{\eps}g_0(x,y)\; ,
\label{e.av2}
\end{align}
where the fast $y$-dynamics is assumed to be ergodic with unique
invariant measure $\mu_x(dy)$ conditioned on the slow variables
$x$. The associated generator is
\begin{align*}
\Lop = \frac{1}{\eps}\Lop_0+\Lop_1\; ,
\end{align*}
with
\begin{align*}
\Lop_0 = g_0(x,y)\partial_y \quad {\rm{and}} \quad \Lop_1 = f_0(x,y)\partial_x\; .
\end{align*}

We consider the case $\Phi(x,y)=x$ and the projection operator
\begin{align}
\label{e.Pav}
(\Proj \omega)(x)= \int \omega(x,y)\,\mu_x(dy)\; .
\end{align}
%where $\mu_x(dy)$ is the invariant measure induced by the fast dynamics (\ref{e.av2}).
We obtain at the first level
\begin{align*}
\dot x &= e^{\Lop t}\Proj\Lop x + n_1\\
&= \langle f_0\rangle + n_1
\end{align*}
with
\begin{align*}
n_1
&= e^{\Lop t}\Q\Lop x
\; ,
\end{align*}
where we introduced $\langle \omega\rangle = (\Proj \omega)(x)$ for
ease of notation. In the limit of infinite time scale separation
$\eps\to 0$ we obtain $n_1=0$ and hence the Mori-Zwanzig equation
reduces to the effective deterministic averaged equation
(\ref{eq.av0}). This is seen by
\begin{align*}
n_1
&= e^{\Lop t}\left(  f_0(x,y)-\langle f_0\rangle \right)\\
&= e^{\LOop \frac{t}{\eps}}\left(  f_0(x,y)-\langle f_0\rangle \right)
+ \int_0^t e^{(t-s)\Lop} \, \Lop_1 \, e^{\LOop \frac{s}{\eps}}\left(  f_0(x,y)-\langle f_0\rangle \right)\, ds \; ,
\end{align*}
where we used the Duhamel-Dyson formula (\ref{eq.Duhamel}). Using the
{\em large deviation principle} whereby deviations of time averages
from the corresponding spatial averages are rare
\citep{MelbourneNicol08}, we argue
\[
\lim_{\eps\to 0}e^{\LOop \frac{t}{\eps}} f_0(x,y) = \langle f_0 \rangle\; ,
\]
and hence obtain
\begin{align*}
\lim_{\eps\to 0}n_1 = 0\; .
\end{align*}
The asymptotic slow dynamics is then summarized as
\begin{align}
d X =F(X)dt\; ,
\label{eq.av0}
\end{align}
with
\begin{align}
\label{eq.F}
%F(x) = \int f_0(x,y)\rhoeq(y|x)\, dy + f_1(x)\; .
F(x) = \int f_0(x,y)\,\mu_x(dy)\; .
\end{align}
The slow dynamics (\ref{eq.av0}) are the well-known deterministic {\em
  averaged} equations. It is well known that on bounded time scales
$\mathcal{O}(1)$ the slow dynamics of the multi-scale system
(\ref{e.av1})-(\ref{e.av2}) is approximated by (\ref{eq.av0}) (see for
example
\cite{VladimirArnold,VerhulstSanders,Givonetal04,Pavliotis08}).

The above exposition is entirely formal. For deterministic dynamical
systems rigorous theory is established in the case when the chaotic
fast dynamics is hyperbolic and the fast dynamics does not depend on the slow dynamics with $g_0=g_0(y)$ by
\cite{Anosov60,Kifer92,Kifer95,Kifer01,Kifer03,Kifer05}. Therein also
stochastic fluctuations around the mean behavior were treated on longer diffusive time scales (see the next Section\ref{sec.homo} on homogenization). An
open problem is how to treat the general case where the slow variables
couple back to the fast chaotic system, i.e. $g_0=g_0(x,y)$, and the
fast dynamics is not hyperbolic. Difficulties occur if the measure
$\mu_x(dy)$ does not depend smoothly on the slow variable $x$ -- this
is, for example, the case when the fast dynamics experiences
bifurcations upon varying $x$. In this case the averaged vector field
may not even be a continuous function of the slow variable and
unique solutions of the averaged equations (\ref{eq.av0}) are not
guaranteed. This non-smoothness of the invariant measure can occur in simple dynamical systems (see \citet{BaladiSmania08}), and may cause problems when trying to apply
linear response theory in climate science.

%%%%%%%%%%%%%

\subsection{Homogenization in the Mori-Zwanzig framework}
\label{sec.homo}
The slow averaged equations (\ref{eq.av0}) are only valid on bounded
time scales of order $\mathcal{O}(1)$ and solutions of the averaged equation (\ref{eq.av0})  will not be close to solutions of the slow variable of the full multi-scale system (\ref{e.av1})-(\ref{e.av2}) on long time scales. We
consider here the case when the averaged drift is small with $\langle
f_0 \rangle = \mathcal{O}(\varepsilon)$. In this case fluctuations
become important. To illustrate how stochasticity and diffusive
behavior arises asymptotically in multi-scale systems on long time
scales, we begin with a simplified version of the general dynamical
system (\ref{e.av1})-(\ref{e.av2}) in which the fast chaotic dynamics
drives the slow dynamics non-multiplicatively and the slow dynamics
does not couple back into the fast dynamics, i.e.
\begin{align}
\dot x &=\frac{1}{\varepsilon} f_0(y) + f_1(x,y)
\nonumber
\\
\dot y &= \frac{1}{\varepsilon^2}g_0(y)\; .
\label{e.homofull}
\end{align}
The fast $y$-dynamics is again assumed to be ergodic with unique
invariant measure $\mu(dy)$. The
associated generator is
\begin{align*}
\Lop = \frac{1}{\eps^2}\Lop_0+\frac{1}{\eps}\Lop_1 + \Lop_2\; ,
\end{align*}
with
\begin{align*}
\Lop_0 = g_0(y)\partial_y
\;, \quad
\Lop_1 = f_0(y)  \partial_x
\quad {\rm{and}} \quad
\Lop_2 = f_1(x,y) \partial_x\; .
\end{align*}
Upon neglecting $(\Proj f_0)(x)=\langle f_0\rangle={\cal{O}}(\eps)$ we
obtain at the first level of the Mori-Zwanzig formalism
\begin{align}
\dot x &= e^{\Lop t}\Proj\Lop x + n_1
\nonumber
\\
&= \langle f_1\rangle + n_1
\label{e.xn1}
\end{align}
with
\begin{align*}
n_1
&= e^{\Lop t}\Q\Lop x\\
&=e^{\Lop t}(\Id-\Proj)\Lop x\\
&=e^{\Lop t}\left(f_1(x,y)-\langle f_1\rangle(x)\right) + \frac{1}{\eps}e^{\Lop_0\frac{t}{\eps^2}}f_0(y)
\; .
\end{align*}
The first term of $n_1$ vanishes in the limit $\eps \to 0$ as part of
the large deviation principle mentioned in the previous section. The
second term gives rise to noise in (\ref{e.xn1}) as can be motivated as
follows. Integrating the second term we obtain
\begin{align*}
\frac{1}{\eps}\int_0^{t}e^{\Lop_0\frac{\tau}{\eps^2}}f_0(y)\, d\tau
=
\eps\int_0^{\frac{t}{\eps^2}}f_0(y(s))\, ds\; .
\end{align*}
For sufficiently chaotic fast dynamics one may evoke the central limit
theorem for $\eps \to 0$ to justify
\begin{align*}
%e^{\Lop_0\frac{t}{\eps^2}}f_0(y) = \dot W_{\tfrac{t}{\eps^2}}\, ,
\frac{1}{\eps}e^{\Lop_0\frac{t}{\eps^2}}f_0(y) = \dot W_t\, ,
%\frac{1}{\eps}\int_0^t e^{\Lop_0\frac{\tau}{\eps^2}}f_0(y)\, d\tau = W_t\, ,
\end{align*}
where $W_t$ is an $n$-dimensional Brownian motion with covariance
matrix $\Sigma$ given by the Green-Kubo type relation
\begin{align*}
\frac{1}{2}\Sigma \Sigma^T &=\int_0^\infty \Proj \left( f_0(y) \, e^{\LOop\frac{t}{\eps^2}} f_0(y)\right)\, dt \; .
\end{align*}
Hence, summarizing, on long time scales $\mathcal{O}(\tfrac{t}{\eps^2})$ the slow
dynamics (\ref{e.xn1}) is given by the {\em homogenized} equation
\begin{align}
d X =F(X)dt + \Sigma \, dW_t\; ,
\label{e.homo}
\end{align}
where the drift coefficient is given by
\begin{align*}
F(x) = \int f_1(x,y)\,\mu(dy)\; .
\end{align*}

\noindent
In the more general case
\begin{align}
\dot x &=\frac{1}{\eps}f_0(x,y) + f_1(x)
\label{e.homo1}
\\
\dot y &= \frac{1}{\eps^2}g_0(y)\; ,
\label{e.homo2}
\end{align}
we expect
\begin{align*}
\eps\int_0^{\tfrac{t}{\eps^2}}e^{\Lop t}f_0(x,y)\, dt
%=
%\eps\int_0^{\frac{dt}{\eps^2}}f_0(x,y(s))\, ds\; ,
\end{align*}
to converge to Brownian motion $W_t$ with variance $\Sigma(x)$. Now
the question arises how to interpret the stochastic integral of
$\Sigma(x)dW_t$.
It is well known that stochastic integrals $\int \Sigma(x)dW_t$ are very sensitive with respect to the approximation of the Brownian motion with the It\^o and the Stratonovich interpretations being two cases. A reader-friendly discussion on the It\^o
versus Stratonovich issue is contained in the book by
\cite{Horsthemke}.
The Wong-Zakai theorem and its extensions
\citep{WongZakai65,IkedaWatanabe} provide general conditions under which
convergence holds with the Stratonovich interpretation for the
stochastic integral.  The rationale behind the Stratonovich
interpretation of the noise in homogenized equations is that here rough noise arises as a limit involving only smooth functions of the
smooth deterministic system. Hence in the limit classical calculus
should prevail implying the Stratonovich interpretation. We remark
that the conditions for the Wong-Zakai theorem are satisfied in the
case of one-dimensional slow variables, but may fail in higher
dimensions. The multi-dimensional homogenized equations associated
with (\ref{e.homo1})-(\ref{e.homo2}) are given by
\begin{align}
d X &= F(X) dt + \Sigma(X) dW_t \;,
\label{e.homog}
\end{align}
where $W_t$ denotes $m$-dimensional Brownian motion and the drift
coefficient is given by
\begin{align*}
F(x) = \int f_1(x,y)\,\mu(dy) + \int_0^\infty ds \int f_0(x,y)\cdot \nabla f_0(x,y(s)) \,\mu(dy) \; ,
\end{align*}
and the diffusion coefficient is defined by
\begin{align*}
\Sigma(X) \Sigma^T(X) &=\int_0^\infty ds\int \left( f_0(y) \otimes f_0(y(s)) + f_0(y(s)) \otimes f_0(y) \right)\, \mu(dy) \; ,
\end{align*}
where the outer product between two vectors is defined as $(a\otimes
b)_{ij} = a_{i}b_{j}$ (see
\citet{PapanicolaouKohler74,IkedaWatanabe,KellyMelbourne14}). It is pertinent to stress that mixing of the fast chaotic flow is not necessary for the stochastic limit systems (\ref{e.homo}) and (\ref{e.homog}) to exist.\\

Homogenisation is well understood in the context of multi-scale
systems where the fast dynamics is stochastic with a unique invariant
density \citep{Khasminsky66,Kurtz73,Papanicolaou76}, see also
\citet{Givonetal04,Pavliotis08}. Rigorous results for diffusive
limits of deterministic dynamical systems have only recently been
obtained
\citep{MelbourneStuart11,GottwaldMelbourne13c,KellyMelbourne14}.
It is pertinent to mention that these rigorous results do not make any assumptions on the mixing properties of the fast chaotic dynamics as assumed in most heuristic homogenization approaches such as, for example, in \cite{Majdaetal06}. These
results are, however, at this stage restricted to the case where the
slow dynamics does not couple back to the fast dynamics,
i.e. $g=g(y)$. The general case $g=g(x,y)$ is still an important
open question for the same reasons as discussed above for the case of
averaging.

When simulating multi-scale systems such as the climate, one uses
discretizations of the continuous-time dynamical
systems. Homogenization results for the resulting multi-scale maps
yield very different results compared to their associated
continuous-time parents. It was shown in \cite{GottwaldMelbourne13c}
that for a one-dimensional slow deterministic dynamics the homogenized
system is neither of the It\^o nor of the Stratonovich type and may 
yield widely different statistics than the limiting continuos
multi-scale system which converges to a stochastic differential
equation with Stratonovich noise.\\

The idea of homogenization was spearheaded in the climate community by
the celebrated {\em MTV} approach. The acronym {\em MTV} stands for the surnames of the authors of the original paper \citet{MTV99}. The main
message learned from homogenisation for developing reduced stochastic
models is the inclusion of correlated additive and multiplicative
noise (CAM) \citep{MajdaEtAl09} rather than simple additive
noise. Homogenization relates this type of noise to the dependency of
the term $f_0(x,y)/\eps$ on both, $y$ and $x$. To achieve stochastic
consistency between the reduced stochastic system and the multi-scale
parent system, the parameters of the stochastic process are estimated
from a priori knowledge of the climatic behavior of the slow
variables such as matching the autocorrelation function
\citep{MTV02}. The MTV strategy has been successfully applied to
an atmospheric barotropic model on the sphere \citep{Franzke05} and a
3-layer quasi-geostrophic model \citep{FranzkeMajda06}. Both models
have realistic atmospheric circulation features and the MTV approach
is able to derive reduced order models which reasonably capture these
features with as little as 10 resolved modes.

In general, the stochastic reduction techniques as described above do
not respect a possible underlying conservation law of the full
multi-scale system. In particular, if the multi-scale dynamics were
Hamiltonian, energy conservation would not be
guaranteed. \citet{DubinkinaFrank07,DubinkinaFrank10} have illustrated
how the overall statistical properties depend on the conservation
properties of a numerical discretization. The stochastic reduced
normal forms by \cite{MajdaEtAl09} which are inspired by
homogenization theory ensure that the nonlinear drift terms respect
energy conservation, but the CAM noise does not impose any constraint
on energy conservation. In energy conserving systems the noise would
need to be projected onto the surface of constant
energy. \cite{FrankGottwald13} have carried out such a homogenization
method for a simplified Lagrangian particle description of the
shallow-water equations. Energy-conserving fast systems have also been considered in \cite{JainEtAl15} where the slowly-varying energy is treated as an additional slow hidden variable. It is at this stage though still unclear
whether conserving certain dynamical quantities (and not others) may
lead to dynamically and statistically inconsistent states.\\

Although the rigorous theory requires an infinite time-scale separation, i.e. $\epsilon \to 0$, it has been observed in numerical simulations that homogenized reduced equations remain reliable reduced models for the slow dynamics even for moderate time-scale separation. This is a familiar situation in asymptotic methods here the range of validity often extends the notion of what is small. It is, nevertheless, important to devise methods which go beyond the assumption of infinite timescale separation.  For systems that can be seen as weakly coupled dynamical systems a recent interesting direction was proposed by Wouters and co-workers \citep{Wouters:2012,Wouters:2013}. Therein the Mori-Zwanzig formalism was combined with linear response theory to provide a closure of the relevant slow dynamics which does not rely on any time-scale separation and which retains some information form the memory kernel.
%It is, nevertheless, important to devise methods which go beyond the assumption of infinite timescale separation. An recent interesting direction was proposed by Wouters and co-workers \citep{Wouters:2012,Wouters:2013}. Therein the Mori-Zwanzig formalism was combined with linear response theory to provide a closure of the relevant slow dynamics which does not rely on any time-scale separation and which retains some information form the memory kernel. It is well known that the very notion of time-scale separation is often flawed in atmosphere and ocean dynamics. Relevant observables often contain slow and fast components. Moreover, large-scale features are not necessarily always slow and small-scale features are not necessarily always fast. For example, fast large-scale inertia gravity waves may be larger in scale than slow mainly quasi-geostrophic extra-tropical storms. Going beyond the concept of time-scale separation would be an immense progress.

%%%%%%%%%%%%%

\subsection{What type of noise?}
\label{sec:whatnoise}

A question which so far has not attracted much attention is what type
of noise one can expect as a limit in multi-scale systems? It is tacitly assumed
in the current body of work on reduced stochastic models that fast
degrees of freedom are parameterized by Brownian motion. The current
justification is the central limit theorem. The central limit theorem
indeed holds for a large class of deterministic dynamical systems (see
\cite{MelbourneNicol05,MelbourneNicol09} for mathematical details). In
particular, the central limit theorem holds for {\em strongly chaotic}
systems\footnote{It is pertinent to mention that strong chaoticity is
  not related to an exponential decay of correlations. See
  \cite{GottwaldMelbourne13,GottwaldMelbourne14} for details and
  definitions of {\em strong} and {\em weak} chaoticity.}.
{\em Weakly chaotic} dynamics for which the central limit theorem does
not hold are characterized by a large degree of intermittency whereby
periods of chaotic dynamics are intermittently disturbed by long
laminar periods of seemingly regular behavior. The central limit
theorem, however, can be modified for weakly chaotic dynamics
\citep{Gouezel04}. This has been used by \citet{GottwaldMelbourne13c}
to show that the limiting noise on the homogenized equations is an
$\alpha$-stable noise (or often called L\'evy process).
%; these noises have a power-law decay of the distribution tails).
These non-Gaussian
processes are characterized by the occurrence of jumps of all sizes and have a probability density function with  algebraically decaying so called fat tails. The power-law decay of the distribution tails implies a non-vanishing probability of large jumps and causes an infinite
variance (see for example \cite{CheckinEtAl08} for an
introduction). This type of noise has been observed in planetary-scale
atmospheric circulation \citep{Viecelli98} as well as in abrupt
millennial scale climate changes during the last ice age in ice-core
data \citep{Ditlevsen99}.

In one dimension the stochastic integrals arising in the reduced
homogenized dynamics are then to be interpreted in the sense of Marcus
integrals (see \cite{Applebaum,ChechkinPavlyukevich14}) which is the
interpretation allowing for the validity of classical calculus akin to
the Stratonovich integral in the case of Brownian motion. It is well
known that the simple occurrence of fat tails may not necessarily
imply an $\alpha$-stable distribution but may as well be associated
with multiplicative Brownian noise. \citet{PenlandEwald08} suggested
therefore to favor Brownian CAM noise over L\'evy noise for practical
purposes. However, these two processes are dynamically entirely
different and, moreover, can be distinguished with relatively easy
diagnostic tools such as the $p$-variation
\citep{HeinImkellerPvalyukevich,BurneckiWeron10,Burnecki}. We believe
it will be an interesting avenue to study how intermittent dynamics,
caused by for example persistent atmospheric pattern such as blocking,
can lead to fat tails in the probability density function of slow
processes such as ocean temperatures using homogenization techniques.

%%%%%%%%%%%%%%%%%%%%%%%%%%%%%%%%%%%%%%%%%%%%%

\section{Data-driven models}
\label{sec-DDM}

The reduction techniques described in the previous section provide a
systematic approach to analytically derive stochastic models for the dynamics of
slow degrees of freedom in the climate system. However, there are
situations where these techniques are not feasible, e.g. because of
the complexity of the underlying model equations, or because of the
absence of a clear scale separation. In such cases, a useful alternative
approach can be to infer stochastic models from data. These data,
usually in the form of time series, can come from observations of the
real climate system, but it can also be useful to infer reduced
stochastic models from data obtained from simulations with
comprehensive numerical models.

The central task in this data-driven approach is one of statistical
inference: one must fit stochastic processes from a suitable class to
the observations at hand. The most commonly used class is formed by
diffusion processes, i.e. models consisting of stochastic differential
equations (SDEs) driven by standard Brownian motion. Other classes
considered in this context include L\'evy processes, discrete
processes (finite-state Markov chains) and Hidden Markov Models
(HMMs).

Let us consider a general $d$-dimensional diffusion process $X(t) \in
\Omega \subseteq \R^d$ with corresponding SDE
\begin{equation}
dX(t) = b(X(t))\, dt+\sigma(X(t)) \, dW(t) \, ,
\label{eq:genSDE}
\end{equation}
in which $W(t)$ is a $d$-dimensional vector of independent Wiener
processes. We define the diffusion coefficient as
\begin{equation}
a(x) : = \sigma(x) \sigma(x)^T \, .
\label{eq:diffcoeff}
\end{equation}
Note that for $d>1$, $b$ is vector-valued and $a$ is
matrix-valued. Furthermore, we focus on situations where the drift
$b(X(t))$ and diffusion $a(X(t))$ do not depend explicitly on
time, but only implicitly through their dependence on $X(t)$.

Inferring the functions $b(X(t))$ and $a(X(t))$ from time series
(observations) of $X(t)$ can be very challenging. A key difficulty is
that the finite-time transition density of the process
(\ref{eq:genSDE}) is in general unknown, i.e. there is no closed-form
expression, in terms of $b$ and $a$,
%for the density of $X(t+\Delta t)$ given $X(t)$.
for the density at time $t+\Delta t$ given the density at time $t$.
%% GAG: regarding the change above: Is this what you meant, Daan?
%% DTC: yes, that's fine
Since observations are usually discrete in time, this is a major problem for inference procedures that rely on the likelihood function.
%However, the observations are usually discrete in
%time, so for inference procedures that rely on the likelihood function
%it is a major problem that the transition density is unknown.

In the simplest situation, $b$ is a linear function of $X(t)$,
$\sigma$ is a constant, and the process is univariate ($d=1$), so that
(\ref{eq:genSDE}) is a scalar Ornstein-Uhlenbeck process (in fact, the
OU process is one of the few diffusion processes for which the
transition density is known). Difficulties arise if $b$ is nonlinear,
$\sigma$ is $X(t)$-dependent (multiplicative noise) or $d>1$ (or a
combination of these). A more detailed discussion of these difficulties can be found in
\cite{gobet2004nonparametric,sorensen2004parametric} and references
therein.

\subsection{Linear Inverse Modeling}

The technique of so-called Linear Inverse Modeling (LIM) is used
frequently in climate science to fit stochastic models with linear
drift and additive noise to data. It has, amongst others,  been used for modeling and
predicting sea surface temperatures in the equatorial Pacific ocean
(e.g. \cite{penland1993prediction,penland1995optimal}) and atmospheric
low-frequency variability (e.g. \cite{winkler2001linear}). Assuming zero mean, the SDE of such a linear model with
additive noise is
\begin{equation}
dX(t) = B \, X(t) \, dt + L \, dW(t) \, ,
\label{eq:multOU}
\end{equation}
where $B$ and $L$ are both $(d \times d)$ constant matrices. This
diffusion process has a Gaussian invariant probability distribution
(provided $B$ is negative definite), and is not a suitable model
for phenomena that are manifestly non-Gaussian. This restriction to the simple
class of Gaussian processes with constant diffusion, however, has the major
advantage that the inference of the matrices $B$ and $L$ from observational
data can be done in a computationally efficient way even for high-dimensional
systems with large $d$.

We define $C(\tau)$ as the lag-$\tau$ covariance matrix of $X(t)$,
i.e. its matrix elements are the expectations
\begin{equation}
C_{ij}(\tau) = \mathbb{E}[ X_i(t+\tau) X_j(t)] \, .
\label{eq:Ctau}
\end{equation}
From (\ref{eq:multOU}) it follows that $C(\tau)$, with any $\tau>0$,
and $C(0)$ are related according to
\begin{equation}
C(\tau) = \exp(B\, \tau) \, C(0) \, .
\label{eq:expBtau}
\end{equation}
If we have time series data available with sampling interval $\Delta
t$, i.e. a set of observations $\{X^\text{obs}(0), X^\text{obs}(\Delta
t), X^\text{obs}(2\Delta t), ..., X^\text{obs} (N\Delta t)\}$, we can
estimate the elements of $C(0)$ and $C(\Delta t)$, assuming ergodicity of the underlying dynamical process, as
\begin{eqnarray}
\hat C_{ij} (0) & = & \frac{1}{N} \, \sum_{n=0}^N X_i^\text{obs}(n\Delta t) X_j^\text{obs}(n\Delta t) \\
\hat C_{ij} (\Delta t) & = & \frac{1}{N} \, \sum_{n=0}^N X_i^\text{obs}(n\Delta t) X_j^\text{obs}((n-1)\Delta t)
\; .
\end{eqnarray}
The estimate of $B$ can then be obtained using (\ref{eq:expBtau}) as
\begin{equation}
\hat B = (\Delta t)^{-1} \, \log [ \hat C(\Delta t) \, (\hat C(0))^{-1} ] \, .
\label{eq:estimB_LIM}
\end{equation}
We note that computing the logarithm of a matrix, as is done in
(\ref{eq:estimB_LIM}), is not entirely trivial due to the
non-uniqueness of the logarithm (see \cite{higham2008functions} for more details).

An estimator for the matrix $L$ can be obtained from the equations for
the second moments (covariances) of the process (\ref{eq:multOU}). Let
$\rho(x,t)$ be the probability density function at time $t$ associated
with (\ref{eq:multOU}). The second moments are
\begin{equation}
\langle x_p \, x_q \rangle_\rho := \int_\Omega dx \, \rho(x,t) x_p \, x_q \, .
\end{equation}
Clearly, the time evolution of the moments is determined by the time
evolution of $\rho$, which is in turn governed by the Fokker-Planck
equation. For the process (\ref{eq:multOU}) it reads
\begin{equation}
\frac{\partial}{\partial t} \, \rho(x,t) = - \, \sum_i \frac{\partial}{\partial x_i} \, (Bx)_i \rho(x,t) +
\frac12 \, \sum_{i,j} A_{ij} \frac{\partial^2}{\partial x_i \, \partial x_j} \, \rho(x,t)\; ,
\label{e.FP}
\end{equation}
where $A=LL^T$.
If we assume that $\rho(x,t)$ and its first spatial derivatives are
zero at the boundary of $\Omega$ for all $t$, it is straightforward to
derive that
\begin{equation}
\partial_t \, \langle x_p \, x_q \rangle_\rho = \sum_j B_{pj} \langle x_j \, x_q \rangle_\rho
+ \sum_j B_{qj} \langle x_p \, x_j \rangle_\rho + A_{pq}
\end{equation}
As $t \to \infty$, $\rho$ tends to the invariant density, so that
$\partial_t \, \langle x_p \, x_q \rangle_\rho \to 0$ and $\langle x_p
\, x_q \rangle_\rho \to C_{pq}(0)$, cf. (\ref{eq:Ctau}). Thus, we have
\begin{equation}
B \, C(0) + C(0) \, B^T + A = 0 \, .
\label{eq:FluctDiss}
\end{equation}
Together with the estimates $\hat B$ and $\hat C(0)$ obtained before,
this relation can be used to determine an estimate $\hat A$ of the matrix $A$. For the final
step, arriving at an estimate of the matrix $L$ that appears in
(\ref{eq:multOU}), one can use a Cholesky decomposition of the estimate
$\hat A$. Because $\hat A$ is symmetric
positive-definite, the Cholesky decomposition is unique \citep{golub2013matrix}.
However, other decompositions are possible and $\hat L$ is
not uniquely defined by $\hat A$ since, if $\hat A = \hat L \, \hat L^T$ then
also $\hat A = \tilde L \, \tilde L^T$ with $\tilde L = \hat L \, Q$ for any orthogonal matrix $Q$, i.e. $Q\,Q^T=1$.
This non-uniqueness reflects the fact that the same diffusive
behavior as described by the Fokker-Planck equation (\ref{e.FP}), can be
generated by different stochastic differential equations
(\ref{eq:multOU}) if they have different $L$ (but the same $L \, L^T$ and $B$).

To summarize, for the LIM procedure one needs to compute estimates for
the two covariance matrices $C(0)$ and $C(\Delta t)$ from the
observations $X(0)$, $X(\Delta t)$, ..., $X(N\Delta t)$. The estimates
for $B$ and $A$ are then obtained using (\ref{eq:estimB_LIM}) and
(\ref{eq:FluctDiss}). The computations are quite straightforward and
can easily be performed for processes in high-dimensional spaces,
e.g. $d=O(10^2)$.
%% GAG: Is O(100) really large in the context of climate? this could be pushed to much higher dimensions, or not?
%% DTC: O(100) is not very large from a climate point of view, but I'd say it is large from a statistical point of view. Regarding even higher dim: I'm not sure what will eventually become the bottleneck as d grows. Not clear to me right now if more data (i.e. larger N) will be needed if d grows.
A drawback is that it applies, as discussed above, to a rather
restrictive class of Gaussian processes with constant diffusion described by (\ref{eq:multOU}). For further details the
interested reader is referred to
\cite{penland1993prediction,penland1995optimal,winkler2001linear}.

\subsection{Inference for general diffusion processes}

Statistical inference for diffusion processes with nonlinear drift
and/or multiplicative noise is much more difficult than for
(\ref{eq:multOU}), in particular in the case of multivariate processes. Several
approaches have been developed and used in the context of
atmosphere-ocean science. It must be mentioned that statistical
inference for diffusions is a relevant tool for a wide range of
applications in physics, chemistry, biology, econometrics and finance, going well beyond the context of climate science which is the focus of this chapter.
Not surprisingly, there exists a large body of literature on
this topic,
%both in the application fields and in the area of mathematical statistics.
both on the theory in mathematical statistics as well as on applications to specific problems.
We do not attempt to give a broad overview
here (see
e.g. \cite{rao1999statistical,sorensen2004parametric,kutoyants2004statistical,bishwal2008parameter}
for such overviews), rather we focus on a few methodologies that are
used in atmosphere-ocean science.

In one approach, the drift and diffusion functions are inferred using
their statistical definitions as conditional first and second moments
of the process increments. Starting again from the general diffusion
process (\ref{eq:genSDE}), the drift $b(x)$ and diffusion $a(x)$ as
defined in (\ref{eq:diffcoeff}) are related to the increments
$X(t+\Delta t)-X(t)$ as follows:
\begin{eqnarray}
b(x) & = & \lim_{\Delta t \downarrow 0} (\Delta t)^{-1} \EE [X(t+\Delta t) - X(t) \, | \, X(t) = x] \, , \\
a(x) & = & \lim_{\Delta t \downarrow 0} (\Delta t)^{-1} \EE [(X(t+\Delta t) - X(t)) \otimes (X(t+\Delta t) - X(t)) \, | \, X(t) = x]\; .
\label{eq:DEdefs}
\end{eqnarray}
By binning the state space, i.e. subdividing  $\Omega$ into
non-overlapping sets $\Omega_k$ (bins), one can use these definitions
to compute estimates for $b$ and $a$ in each bin from the observations
$X^\text{obs}(t)$. This approach was proposed in the physics community
in \cite{siegert1998analysis,friedrich2000extracting}, and was used to
analyze atmospheric datasets in
e.g. \cite{sura2003stochastic,berner2005linking}. This approach is very general and does not assume a specific functional form for the drift or diffusion, so
it can be applied to processes involving nonlinear drift and non-constant
diffusion. However, its practical use is limited to low-dimensional
processes because the number of bins grows exponentially in
$d$. Therefore the amount of data needed to obtain statistically meaningful estimates in each
bins also grows exponentially with $d$.
%% GAG: Is the above statement correct? It was "grows very rapidly' before
%% DTC: seems fine to me. If the number of bins grows exponentially, the amount of data must grow exp as well (otherwise many bins will be empty)
Another difficulty is that the
estimators based on (\ref{eq:DEdefs}) rely on $\Delta t$ being
small. This becomes a problem if the observations are not
generated by a $d$-dimensional diffusion process
but rather if the underlying dynamics of the observed system is of a deterministic chaotic nature or if the observations are given as a projection of a higher-dimensional process. For example, the limit of the diffusion $a(x)$ becomes zero for $\Delta t\to 0$ in deterministic systems. As for a projected process, this will generally be non-Markov, so the result of fitting a Markov process to it will depend on the choice of $\Delta t$.
%% GAG: Why is there are problem with the sampling time, when the observations are projections form a higher-dimensional stochastic system?
%% DTC: then the observations will generally be non-Markov (due to projection)
The issue of the choice of the sampling time and possible biases of the estimation of drift $b(x)$ and diffusion $a(x)$ for too small or too large observation intervals will be discussed in more detail in Section\ref{sec.IMD}.
%
%(e.g. because the
%observations have an underlying deterministic, chaotic process, or
%because they are a projection of a higher-dimensional stochastic
%process), so that the small $\Delta t$ limit can give large biases in
%the estimates of $b$ and $a$. Using larger $\Delta t$ can also lead to
%large bias, due to the fact that the small $\Delta t$ assumption no
%longer holds. This issue is discussed in more detail later.

The methodology proposed in \cite{Crommelin06,crommelin2011diffusion}
(see also \cite{gobet2004nonparametric})  is partly motivated by the
need to overcome the small $\Delta t$ limit without introducing time
discretization errors as in the estimators based on
(\ref{eq:DEdefs}). At the core of this method lies the relationship
between the conditional expectation operator denoted $P_{\Delta t}$
and the diffusion operator (or backward Fokker-Planck operator)
denoted $\mathcal{L}$. For suitable functions $h(x)$, we define the
former as
\begin{equation}
(P_{\Delta t}\, h)(x) = \EE [h(X(\Delta t)) \, | \, X(0)=x] \, .
\end{equation}
The diffusion generator is
\begin{equation}
\mathcal{L} = \sum_{i=1}^d b_i(x) \partial_i + \tfrac12 \sum_{i,j=1}^d a_{ij}(x) \partial_i \partial_j \, ,
\end{equation}
where $\partial_i$ is shorthand notation for $\partial / \partial
x_i$ (cf (\ref{e.gen})). For diffusion processes, $\mathcal{L}$ is the generator of the
semigroup of operators $P_{\Delta t}$ with $\Delta t \geq 0$
\begin{equation}
(\mathcal{L} \, h)(x) = \lim_{\Delta t \downarrow 0} (\Delta t)^{-1} [(P_{\Delta t} \, h)(x) - h(x)]\; ,
\end{equation}
and thus
\begin{equation}
P_{\Delta t} = \exp (\Delta t \, \mathcal{L}) \, .
\end{equation}
This implies that the eigenfunction-eigenvalue pairs of $P_{\Delta t}$
and $\mathcal{L}$ are closely related, and we identify
\begin{equation}
P_{\Delta t} \, \phi = \Lambda \phi \,\,\,\, \Leftrightarrow \,\,\,\, \mathcal{L} \phi = \lambda \phi \,\,\,\, \text{with} \,\,\,\, \Lambda = \exp(\lambda \, \Delta t) \; .
\label{eq:PLeig}
\end{equation}
Note that this relation is exact and holds for all $\Delta t \geq 0$.
%there is no approximation or assumption of small $\Delta t$
%involved.
A similar relation holds for the adjoints in
$L^2(\Omega,dx)$ of $P_{\Delta t}$ and $\mathcal{L}$ and their
eigenpairs, see \cite{crommelin2011diffusion}.

These relations can be used to estimate the drift and diffusion
functions $b$ and $a$ that determine $\mathcal{L}$ in the following
way: from discrete-in-time observations with sampling interval $\Delta
t$ we can infer (a Galerkin approximation to) the operator $P_{\Delta
  t}$. Denoting the eigenpairs of this estimated operator by $(\hat
\phi_k, \hat \Lambda_k)$ with index $k$ (ordered by decreasing $|\hat
\Lambda_k|$), we compute $\hat \lambda_k = (\Delta t)^{-1} \log \hat
\Lambda_k$, cf. (\ref{eq:PLeig}). From the (leading) estimated
eigenpairs $(\hat \phi_k, \hat \lambda_k)$ we can compute estimates of
$b$ and $a$ by minimizing the residuals $r_k := \mathcal{L} \hat
\phi_k - \hat \phi_k \lambda_k$, in a suitable way, under variation of
$b$ and $a$.

The minimization problem can be formulated with various cost
functions. One example is  a sum of squared norms $\sum_k \alpha_k \|
r_k \|^2$ with weights $\alpha_k$. Another possibility is $\sum_{k,l}
| \langle r_k, \omega_l \rangle |^2$, where $\langle r_k, \omega_l
\rangle$  denotes $r_k$ integrated against suitable test functions
$\omega_l(x)$. We refer to \cite{crommelin2011diffusion} for more
details. Here we only mention that this method can be used for
parametric as well as for non-parametric estimation of $b$ and $a$, and
that the minimization problem is often of convex quadratic form.

This approach has the advantage that it can be used for general
diffusions, and is not dependent on any small $\Delta t$ approximation. It was used in
\cite{thompson2014parametric} for estimating parameters in a
2-dimensional stochastic model for sea surface winds. Notwithstanding,
this method is also limited to low-dimensional processes and since the estimation of the operator $P_{\Delta t}$ from observations becomes impractical for higher dimensional systems.
%% GAG: I deleted the 'Galerkin" because even if one estimates the operator via binning, the estimation becomes impractical as bins grow expoenntially. OK?
%% DTC: yes, that's fine. In fact, binning can also be seen as Galerkin approximation, but we can leave such details out.
%%
%becomes impractical when t too the practical use is limited to low-dimensional
%processes, because it involves estimation of a Galerkin approximation
%to the operator $P_{\Delta t}$.

In \cite{sitz2002estimation}, it is proposed to use a nonlinear
extension of the well-known Kalman filter, the so-called unscented
Kalman filter, for parameter estimation of a diffusion process with
observation noise. This method is used in \cite{kwasniok2009deriving}
to estimate the parameters of a nonlinear drift function in a
1-dimensional model for glacial-interglacial transitions. The data are
provided by an ice-core record from Greenland. This method can handle
nonlinear drift, however it does not provide a way to estimate the
diffusion coefficient. The diffusion must be estimated by a different
method, and is used as input for the unscented Kalman filter
approach.

Finally, we mention here the recent work presented in
\cite{peavoy2015systematic}, where a Bayesian framework is developed
for parameter estimation using Markov Chain Monte Carlo (MCMC)
methods, which is, in principle, applicable to high-dimensional systems. In this work, the structural form of the diffusion process is
motivated by the stochastic mode reduction (MTV) methods for climate
models discussed in the previous section. It
builds on recent advances
(e.g. \cite{chib2004likelihood,golightly2008bayesian}) in Bayesian
inference and MCMC methods for diffusion processes, by imposing physical constraints such as global stability.
%The used MCMC
%scheme was first developed by \cite{golightly2008bayesian}.
This methodology
can handle nonlinear drifts and non-constant diffusions, and is
demonstrated on examples with dimensions of the state vector with
$d=1$ and $d=2$ in \cite{peavoy2015systematic}, but can be extended to higher dimensional problems.
%The novelty of this scheme is that it
%uses physical constraints such as global stability to make the parameter
%estimation more efficient. Without this constraint about 40\% of the
%parameter estimates would lead to physically unstable solutions.

\subsection{Inference from multi-scale data}
\label{sec.IMD}
Stochastic models are often used as coarse-grained models for
phenomena that are generated by a complex dynamical system with many
scales. An example is the model considered in
\cite{kwasniok2009deriving}, where a diffusion process is inferred
from Greenland ice-core data ($\delta^{18} \, O$ values) going back
120 000 years. These $\delta^{18} \, O$  data are a proxy for northern
hemisphere (NH) temperatures in the past, whose dynamics are generated
by the full climate system with its many temporal scales. Thus, a
scalar stochastic model for the dynamics of NH temperature (or
$\delta^{18} \, O$ values) over timescales of centuries and longer is
inevitably a coarse-grained model. Such a model is aimed at capturing
the correct long timescale behavior but not necessarily the dynamics
on short timescales.

An important question in this context is whether statistical inference
from data of a multi scale system will yield an accurate representation
of the long timescale dynamics, or whether it will result in a model
that mainly reflects the short timescale behavior.
%It is shown in \cite{Pavliotis07} that inference from data with a too short sampling interval can give strongly biased estimates, so that the resulting diffusion processes are poor models for the coarse-grained dynamics.
%In \cite{Pavliotis07}, multi scale diffusion processes are
%considered that have a well-defined coarse-grained diffusion process
%in the limit of large scale separation, obtained by averaging or
%homogenization techniques. However, inferring the coarse-grained
%process from data of the full multi scale process can give results that
%are very different from the analytically derived coarse-grained
%process. As already mentioned, this can occur if the sampling interval
%$\Delta t$ is too short.
In \cite{Pavliotis07}, multi scale diffusion processes are
considered which possess a well-defined coarse-grained diffusion process
in the limit of large scale separation, obtained by averaging or
homogenization techniques as described in Sections~\ref{sec.av} and \ref{sec.homo}. Inferring the coarse-grained
process from data of the full multi scale process, however, was shown to yield results that
are very different from the analytically derived coarse-grained
process,
if the sampling interval $\Delta t$ is too short.
%As already mentioned, this can occur if the sampling interval
%$\Delta t$ is too short.
For estimation methods that rely on small $\Delta t$ (e.g. the
estimators in (\ref{eq:DEdefs})), this can be particularly
bothersome. These methods can be caught between $\Delta t$ being too
small (giving the "multi scale bias" discussed above) and $\Delta t$
being too large (giving bias due to the time discretization error of
the estimator). \cite{MitchellGottwald12b} show that with the estimators
(\ref{eq:DEdefs}), one typically obtains a linear drift term if $\Delta t$
is too large, even if it should be nonlinear.
We refer to
\cite{Pavliotis07,papavasiliou2009maximum,crommelin2011diffusion,MitchellGottwald12b,azencott2013sub}
for a more detailed discussion of these issues.

As a final remark, we mention that the dependency of estimation
results on the sampling interval has been noted in atmosphere-ocean
applications as well. In \cite{penland1995optimal}, the so-called
"tau-test" is introduced to test if the results from the LIM method
are independent of the sampling interval (denoted $\tau$ in that
paper, rather than $\Delta t$).
%\cite{penland1995optimal} argue there
%may be several causes if they do depend on the sampling interval (one
%being that nonlinear dynamics are important, so that the functional
%form of equation (\ref{eq:multOU}) is inadequate).
\cite{penland1995optimal} argue that a dependency on the sampling interval points towards a possible inadequacy of the functional
form of equation (\ref{eq:multOU}), and to, for example, possible nonlinearity of the underlying dynamics.
In \cite{berner2005linking}, the
sampling interval dependence of results obtained with the estimators
(\ref{eq:DEdefs}) is demonstrated numerically.
%% GAG: Daan, can you present/summarize some of the results she found? What particular interesting/useful dependency was found?
%% DTC: ok, here goes:
Berner reports that sampling intervals between 1 and 6 days are suitable for inferring a low-order stochastic model for planetary wave behavior from GCM time series data. However, for choosing the sampling interval "there
seems to be a trade-off between reproducing the non-Gaussianities in the PDF versus capturing the temporal
aspects of planetary wave behavior" (quoting \cite{berner2005linking}). A shorter $\Delta t$ gives a better reproduction of the PDF, whereas with a longer $\Delta t$ the temporal decay of correlations is better captured.

Also relevant in this
context is the study by \cite{DelSole00}, who investigates to what extent
stochastic models are able to capture the statistics of deterministic
dynamical systems. He notes, and analyzes in considerable detail, the sampling interval dependency
when fitting stochastic models to deterministic dynamical systems. A key observation
by DelSole is that the shape of the normalized autocorrelation functions (ACFs) of deterministic and stochastic (Markov) models differ at short time lags $(\Delta t)$. For deterministic models, the ACF must be of the form $1-\gamma \, (\Delta t)^2$ with $\gamma$ a positive constant if $\Delta t$ is small, and thus the derivative of the ACF with respect to $\Delta t$ vanishes as $\Delta t \downarrow 0$.
By contrast, the ACF of frequently used Markov stochastic processes is of the form $\exp(-\beta |\Delta t|)$, with decay rate $\beta > 0$, for small $\Delta t$. Its derivative does not vanish but tends to $-\beta$ as $\Delta t \downarrow 0$. One can fit a Markov stochastic process so that its ACF intersects the ACF of the deterministic system at a chosen lag $\Delta t^*$, but the two ACFs will not coincide at other lags. Thus, the fitted stochastic process will depend on the chosen lag.
%% GAG: Daan, can you present/summarize some of the results he found? What particular interesting/useful dependency was found?
%% DTC: done, see above.

\subsection{Beyond diffusion processes}

Diffusion processes driven by standard Wiener processes, such as
(\ref{eq:genSDE}), are not the only type of stochastic process used
for modeling (aspects of) the climate system. A generalization is the
class of diffusions driven by L\'evy processes, as already discussed in section
\ref{sec:whatnoise}.
%There, the issue of distinguishing between L\'evy noise and Brownian noise is considered.
Inference for L\'evy processes is an active area
of research in mathematical statistics
(e.g. \cite{jongbloed2005nonparametric}), however it has not been used
much for climate applications yet. Some exceptions, already mentioned in \ref{sec:whatnoise}, are by
\citet{Viecelli98,Ditlevsen99}.

A class of stochastic processes that has been used more widely in atmosphere-ocean science is
that of finite-state Markov chains. These have been employed, for example, to study the regime behavior in the atmosphere \citep{spekat1983grosswetter,mo1987statistics,crommelin2004observed}. Therein the
transitions between a finite number of preferred states of the large-scale atmospheric
flow (so-called regimes) are modeled as a Markov chain. If a time series of such
finite-state dynamics is given, it is straightforward to estimate the
elements of the transition probability matrix (or stochastic matrix)
that defines the Markov chain.

Another application where finite-state Markov chains have been used
is stochastic parameterization of atmospheric convection, see e.g.
\citet{khouider2003coarse,khouider2010stochastic,dorrestijn13,
dorrestijn15a,dorrestijn15b,gottwaldpetersdavies2015}). In this approach,
the range of convective activity of any atmospheric model column is discretized into a few
distinct states. The transitions between these states as time evolves can then
be modeled as a Markov chain, with transition probabilities that depend on
the large-scale state of the atmosphere. The discretization has been carried
out in several ways, e.g. \citet{dorrestijn13} discretize the vertical
turbulent fluxes of heat and moisture using a clustering method, whereas
\citet{khouider2010stochastic} and \citet{dorrestijn15a} employ a small
number of cloud states for discretization. In \citet{gottwaldpetersdavies2015},
it is the convective area fraction that is discretized. Furthermore, the
transition probabilities are obtained in different ways, using either
physical intuition (e.g \citet{khouider2003coarse,khouider2010stochastic})
or statistical inference (e.g. \citet{dorrestijn13,
dorrestijn15a,dorrestijn15b,gottwaldpetersdavies2015}).

In \cite{pasmanter2003cyclic}, Markov chains are used for studying
ENSO predictability. The influence of the seasonal cycle is accounted
for by employing so-called cyclic Markov chains: twelve different
stochastic matrices $\mathbf{m}_i$ ($i=1, ..., 12$) are constructed
(or in fact estimated), each specifying the transition probabilities
of the various states from month $i$ to month $i+1$. The transition
probabilities from month $i$ to the same month $i$ one year later is
then given by the product $\mathbf{M}_i = \mathbf{m}_{i-1}
\mathbf{m}_{i-2} \cdots \mathbf{m}_1 \mathbf{m}_{12} \mathbf{m}_{11} \cdots \mathbf{m}_i$.
The $\mathbf{M}_i$ are again
stochastic matrices. Furthermore, \cite{pasmanter2003cyclic} construct
their Markov chains by equipartition of the data, resulting in
transition probability matrices that are in fact doubly stochastic
matrices, satisfying both $\sum_k \mathbf{m}_i (k,l) = 1 \,\, \forall
l$ and $\sum_l \mathbf{m}_i (k,l) = 1 \,\, \forall k$. This
facilitates the analysis of their model in terms of Floquet theory and
information loss properties. Some of these concepts are also used in
\cite{crommelin2004observed}.

The Markov chains as described above have finite state spaces. They
are frequently used to model the dynamics of continuous quantities,
requiring the discretization of these quantities. In the framework of
Hidden Markov Models (HMMs), this discretization is no longer
needed. HMMs still employ a finite state Markov chain, however the
observed quantities or time series are assumed to be generated by
another process whose properties depend on the Markov chain
state. That other process may be continuous or discrete in time. An example is
the case where the observations are independent draws from a normal
distribution. In this case, if we denote the observations by $Y_t$, we
have $Y_t \sim \mathcal{N} (\mu_k,\sigma^2_k)$. The values of the mean
$\mu_k$ and variance $\sigma^2_k$ change over time, they can take on a
finite number of values $\{ \mu_1, ..., \mu_K \}$ and $\{ \sigma^2_1,
..., \sigma^2_K \}$. The index $k$ changes randomly in time in
accordance with a Markov chain with $(K \times K)$ stochastic
matrix. The value of $k$ is unobserved (or hidden).

In the context of atmosphere-ocean science, HMMs have been used to
model precipitation (e.g. \cite{zucchini1991hidden}) as well as
dynamics of large-scale atmospheric flow (e.g. \cite{Franzke08}). The
statistical inference for HMMs with normal distributions as output is
tractable through the expectation-maximization algorithm, see
\cite{Franzke08} for more details and references.

%It must be emphasized that the overview of data-driven methods
%presented here is by no means exhaustive. We have mainly focused on
%diffusion processes and Markov chains here, leaving out
%e.g. time series methods (ARMA models etc) for the sake of brevity. We
%conclude by pointing out two more lines of research relevant in the
%context of this section. Egger (e.g. \cite{egger2001master}) has
%employed master equations inferred from time series to study
%atmospheric phenomena, an approach closely related to Markov chain
%modeling. Furthermore, Horenko (e.g. \cite{horenko2010identification})
%has developed techniques to deal with non-stationarity in timeseries
%data, an issue left out of consideration in most studies on
%data-driven approaches.

We conclude by pointing out two more lines of research relevant in the
context of this section. Egger (e.g. \cite{egger2001master}) has
employed master equations inferred from time series to study
atmospheric phenomena, an approach closely related to Markov chain
modeling. Furthermore, Horenko (e.g. \cite{horenko2010identification})
has developed techniques to deal with non-stationarity in time series
data, an issue left out of consideration in most studies on
data-driven approaches.
It must be emphasized that the overview of data-driven methods
presented here is by no means exhaustive. We have mainly focused on
diffusion processes and Markov chains here, leaving out
e.g. time series methods (ARMA models etc) for the sake of brevity.

%%%%%%%%%%%%%%%%%%%%%%%%%%%%%%%%%%%%%%%%%%%%%

\section{Outlook}
In this chapter we have described current approaches to either systematically
derive reduced order stochastic climate models or to extract the stochastic dynamics from
observed data. The two approaches of data-driven models and analytic physics-based models are complementary. In practice, in the future both approaches should be
combined where the Mori-Zwanzig formalism provides the functional form for
model fitting to observed data. This will enable us to fit more complex models
with the currently available amount of data. \citet{MajdaHarlim13},
\citet{peavoy2015systematic} and \citet{KondrashovEtAl15} put forward such
physics constrained approaches which are based on energy conservation
and global stability. Such reduced order models are used in many
practical applications like long-range climate forecasts (e.g. El
Ni\~no-Southern Oscillation (ENSO)) or weather and climate catastrophe
modeling \citep{Born:2006}.

Another area where stochastic approaches are actively investigated is that of
parameterizations, i.e. simplified representations of spatially localized,
small-scale physical processes such as atmospheric convection.
The need for stochastic parameterizations in
complex numerical weather and climate prediction models becomes ever
more clearer. A recent study by \citet{DawsonPalmer14} showed that the
European Centre for Medium Range Weather Forecasts (ECMWF) model with a
stochastic physics scheme performs as well as a
purely deterministic model version at a much higher horizontal
resolution. Hence, stochastic weather and climate models offer the
potential of achieving more accurate simulations at a lower
computational expense. However, most of the current stochastic
parameterization approaches are mainly ad hoc schemes
\citep{Shutts04,Shutts05,Berner09,Franzke15}. There is a pressing need
to base these stochastic parameterizations on a more sound
mathematical and physical footing. Current systematic approaches for
doing this include Frederiksen et al. (chapter in this book and
references therein), \citet{khouider2003coarse,
crommelin2008subgrid,plant2008stochastic,khouider2010stochastic,Wouters:2012,Wouters:2013,dolaptchiev2013stochastic,grooms2013efficient},
but more fundamental work in this area is clearly needed. See also the
review by \citet{Franzke15}.
Moreover, the Mori-Zwanzig formalism
shows that such parameterization schemes might have to take memory
effects into account. Most schemes currently do not account for this,
and the issue of memory effects has hardly been explored yet in the context of
parameterizations. Some exceptions are \citet{crommelin2008subgrid,
verheulcrommelin2015,gottwaldpetersdavies2015,chorinlu2015}.

Weather and climate prediction models use high-performance
computers. These computing systems are expected to reach soon limits
regarding energy use and heat production. These issues led to attempts
to use imprecise computational techniques or stochastic processors
\citep{Palmer:2014}. It is thought, that these techniques can be
carried out on less energy consuming computer systems. However, the
use of stochastic processors also needs a firm mathematical
underpinning since the implementation of the stochastic noise produced
by the processors needs to be appropriate. This is another area for
future research.

Stochastic approaches are also important for the analysis of observed
data and the understanding of their characteristics. In particular, the
detection and attribution of forced trends is an important current research
topic. For instance, there is currently a debate going on in the climate
science community whether climate variability is long-range dependent (LRD) or
whether it is better better described as short-range dependent (SRD). LRD
systems are able to produce more persistent stochastic trends than SRD
systems. So, if the climate system is LRD but is investigated with SRD methods
then one is likely to mistake a stochastic LRD trend for a significant
externally forced trend. Hence, the detection and attribution of external
trends is hampered in LRD systems. See chapters by Bunde et al. and Watkins
for more details. This topic is critically discussed in the contemporary
climate literature and many climate scientists are sceptical about whether the
climate system is LRD because there is a lack of physical mechanisms
explaining the LRD characteristic \citep{Franzke:2015}. The Mori-Zwanzig
formalism provides an explanation how memory potentially arises in the climate
system. However, whether this can explain LRD needs further research.

\mbox{}\\

\textbf{Acknowledgments}\\
We would like to thank Jeroen Wouters for a careful reading of the
manuscript. GAG acknowledges funding from the Australian Research Council. The
research of DTC is supported by the Netherlands Organisation of Scientific
Research (NWO) through a Vidi grant. CF was supported by the German Research
Foundation (DFG) through the cluster of excellence CliSAP (EXC177).

%%%%%%%%%%%%%%%%%%%%%%%%%%%%%%%%%%%%%%%%%%%%%

%\bibliographystyle{natbib}
%\bibliography{bibliography}

%%%%%%%%%%%%%%%%%%%%%%%%%%%%%%%%%%%%%%%%%%%%%

\end{document}